\documentclass[11pt,a4paper]{article}
\usepackage[T1]{fontenc}
\usepackage[utf8]{inputenc}
\usepackage{lmodern}
\usepackage{jheppub}
\usepackage{mathtools}
\usepackage{booktabs}
\usepackage{xurl,needspace,placeins}
\allowdisplaybreaks[2]
\makeatletter
\gdef\@fpheader{}
\makeatother
\hypersetup{hypertexnames=false,pdftitle={Quantum hydrodynamics from Nambu's unpublished fluid notes},
  pdfsubject={Fluid geometry, quantum channels, and coherence between opposite momenta},
  pdfauthor={Kazuki Ikeda}}
\newcommand{\scinum}[2]{\mbox{\ensuremath{#1\times10^{#2}}}}
\newcommand{\dd}{\mathop{}\!\mathrm{d}}
\newcommand{\Tr}{\operatorname{Tr}}
\newcommand{\Var}{\operatorname{Var}}
\newcommand{\id}{\operatorname{id}}
\newcommand{\cof}{\operatorname{cof}}
\newcommand{\ket}[1]{\lvert #1\rangle}
\newcommand{\bra}[1]{\langle #1\rvert}

\let\ArticleTableOfContents\tableofcontents
\renewcommand{\tableofcontents}{\begingroup\renewcommand{\baselinestretch}{1.0}\small\ArticleTableOfContents\endgroup}
\renewcommand{\afterTocRuleSpace}{\clearpage}
\title{Quantum hydrodynamics from Nambu's unpublished fluid notes}
\author[a,b]{Kazuki Ikeda}
\affiliation[a]{Department of Physics, University of Massachusetts Boston, Boston, MA 02125, USA}
\affiliation[b]{Center for Nuclear Theory, Department of Physics and Astronomy, Stony Brook University, Stony Brook, NY 11794-3800, USA}
\emailAdd{kazuki.ikeda@umb.edu}
\abstract{We develop a quantum hydrodynamic theory based on Nambu's unpublished manuscript \emph{A New Look at Fluid Dynamics}. The kinetic action induces a metric on a family of volume-preserving configurations. For a cyclic coordinate, the mechanical connection enters the Hamiltonian through a coupling odd in the conserved momentum. We derive a geometric criterion for exact preservation of coherence between opposite momenta and relate short-time decoherence to the variance of the momentum-dependent coupling. Two shear flows realize an exact noiseless subsystem encoded in the sign of momentum. Additional deformations cancel the initial Euler residual and generate nonzero connection curvature, which obstructs local elimination of the mixed metric components. In a three-coordinate model, we construct an initial amplitude annihilated by the leading-order coupling. In our numerical examples at equal mean energy, the resulting adapted state has a short-time decoherence coefficient more than three orders of magnitude smaller than that of the product state. At later times, the product state can retain greater coherence. These results relate the kinetic geometry of incompressible fluids to quantum coherence.}

\begin{document}
\maketitle
\raggedbottom

\section{Introduction}
\label{sec:introduction}
In his unpublished manuscript \emph{A New Look at Fluid Dynamics}, Yoichiro Nambu considered the possibility of quantizing fluid motion through mappings of space~\cite{note}. Using a stream function in two dimensions and pairs of generating functions in three dimensions, he described volume-preserving flows and examined their compatibility with the Euler equation. We use this formulation to quantize fluid deformations and study the coherence of one observed mode.

The kinetic geometry of incompressible flow underlies Arnold's formulation of ideal-fluid motion~\cite{arnold} and later structure-preserving discretizations~\cite{pavlov}. Related geometric formulations employ the Nambu bracket~\cite{nambu1973,takhtajan} and canonical Clebsch variables~\cite{yoshida}.

Effective field theories of perfect fluids can be formulated in terms of comoving scalar fields with volume-preserving internal symmetries. Relabeling symmetry constrains the low-energy action and its derivative expansion~\cite{dubovsky2012}. Quantization raises infrared questions associated with vortex degrees of freedom~\cite{endlich}. Dersy, Khmelnitsky and Rattazzi developed a quantum description of the two-dimensional incompressible fluid using finite-dimensional approximations to the area-preserving diffeomorphism group and studied its vorton excitations~\cite{dersy2024}. Goldberger and Tadi\'c analyzed quantum perfect fluid correlators in semiclassical Gaussian initial states, including transverse contributions to the stress-tensor response~\cite{goldbergertadic2026}.

Geometric couplings also arise in nuclear collective motion. In the Bohr description of quadrupole vibrations and rotations, deformation-dependent inertia defines a metric for quantization~\cite{prochniak2009}. Microscopic collective Hamiltonians use inertial functions and a potential obtained from self-consistent mean-field calculations~\cite{niksic2009}, and effective field theories describe excitations at different energy scales~\cite{coelloperez2025}. In deformed odd-mass nuclei, gauge potentials couple a valence nucleon to the rotating core and relate Coriolis effects to Berry phases~\cite{papenbrock2020}. These developments motivate the study of quantum coherence in systems with deformation-dependent inertia.

We consider three-dimensional incompressible flow with finitely many collective coordinates. Compositions of flows generated by pairs of scalar functions define a family of volume-preserving configurations. The induced kinetic metric governs their classical dynamics. For the quantum theory, we adopt the Laplace--Beltrami Hamiltonian on the resulting configuration space~\cite{dewitt}. The observed and unobserved modes evolve under the total Hamiltonian, and tracing out the latter yields the reduced density operator. We examine the change in coherence when additional coordinates improve the approximation to the Euler equation.

When the observed coordinate $q$ is cyclic, the kinetic metric defines a mechanical connection~\cite{marsdenphases,littlejohn} with local connection form $A$. The part of the Hamiltonian odd in the conserved momentum is $pB$, where $B$ depends on $A$ and the reduced metric of the unobserved coordinates. When $A=0$ in the chosen coordinates, the Hamiltonian is even in the conserved momentum, whose sign defines a noiseless subsystem~\cite{knill,lidar}. The accessible observable algebras specify the subsystem decomposition~\cite{zanardi}. Nonzero curvature obstructs local elimination of $A$ by redefining the cyclic coordinate. The variance of the momentum-dependent coupling appears in the quadratic term of the coherence loss.

Two shear flows realize a noiseless subsystem encoded in the sign of momentum. Adding two generating flows cancels the initial Euler residual and induces nonzero connection curvature. Restricting their parameters to a single coordinate gives a three-coordinate model in which we suppress short-time decoherence using initial amplitudes annihilated by the leading-order coupling. In numerical comparisons at equal mean energy, the adapted state has a short-time decoherence coefficient more than three orders of magnitude smaller than that of the product state. At later times, the product state can retain greater coherence. For a given momentum magnitude, the conditional overlap has the form of a Loschmidt echo amplitude~\cite{gorin}.

Section~\ref{sec:construction} derives the Hamiltonian and coherence criterion. Sections~\ref{sec:two} and \ref{sec:four} examine the two-shear model and its extensions. Section~\ref{sec:numerical} compares the classical trajectories and quantum coherence. Supplementary derivations, numerical methods, and convergence proofs appear in the appendices.

\Needspace{12\baselineskip}
\section{Kinetic geometry and quantization}
\label{sec:construction}
\subsection{Volume-preserving configurations}
\label{sec:configurations}
Let $\mathbf a$ label the fluid elements on a fixed periodic domain $M$, and let $\mathbf X(\mathbf a,t)$ be their positions. For a reference mass density $\rho_0>0$ and an external potential energy density $U$, we write Nambu's constrained action in continuum notation as~\cite[p.~2]{note}
\begin{equation}
 S=\int\dd t\int_M\dd^3a
 \left[\frac{\rho_0}{2}|\partial_t\mathbf X|^2-U(\mathbf X)
 +\lambda(\mathbf a,t)(J_{\mathbf X}-1)\right].
 \label{eq:action}
\end{equation}
Here $J_{\mathbf X}=\det D_{\mathbf a}\mathbf X$. Variation with respect to $\lambda$ imposes $J_{\mathbf X}=1$, and variation with respect to $\mathbf X$ yields the incompressible Euler equation
\begin{equation}
 \rho_0(\partial_t\mathbf v+\mathbf v\cdot\nabla\mathbf v)
 =-\nabla U-\nabla P,\qquad \nabla\cdot\mathbf v=0,
 \label{eq:euler}
\end{equation}
where $\mathbf v=\partial_t\mathbf X\circ\mathbf X^{-1}$ and $P=\lambda\circ\mathbf X^{-1}$ is the fluid pressure. The derivation and its relation to Nambu's notation are presented in Appendix~\ref{app:geometry}.

In Nambu's description, pairs of scalar functions generate three-dimensional incompressible motion~\cite[p.~4]{note}. Choose $m$ smooth pairs $(\phi_r,\chi_r)$ and their flows,
\begin{equation}
 \mathbf w_r=\nabla\phi_r\times\nabla\chi_r,\qquad
 \partial_s\Phi_r^s=\mathbf w_r\circ\Phi_r^s,\quad \Phi_r^0=\id,
 \label{eq:generators}
\end{equation}
with the prescribed boundary conditions and parameter values within the domains of the flows. Each vector field is divergence-free and tangent to the common level sets of the scalar functions. Composing the flows defines a family parametrized by collective coordinates $\mathbf q=(q^1,\ldots,q^m)$,
\begin{equation}
 \mathbf X_{\mathbf q}=\Phi_m^{q^m}\circ\cdots\circ\Phi_1^{q^1},
 \qquad J_{\mathbf X_{\mathbf q}}=1.
 \label{eq:composition}
\end{equation}
With $\partial_r=\partial/\partial q^r$, the Eulerian velocity is
\begin{equation}
 \mathbf v
 =\sum_r\dot q^r\,\partial_r\mathbf X_{\mathbf q}\circ\mathbf X_{\mathbf q}^{-1}.
 \label{eq:collectivevelocity}
\end{equation}
\subsection{Kinetic metric and quantum dynamics}
\label{sec:reduced}
We henceforth set $U=0$. Restricting the action to Eq.~\eqref{eq:composition} yields
\begin{align}
 L&=\frac12G_{rs}(\mathbf q)\dot q^r\dot q^s,
 &G_{rs}(\mathbf q)&=\rho_0\int_M
 \partial_r\mathbf X_{\mathbf q}\cdot\partial_s\mathbf X_{\mathbf q}\,\dd^3a,
 \label{eq:metric}\\
 p_r&=G_{rs}\dot q^s,
 &H&=\frac12p_rG^{rs}(\mathbf q)p_s.
 \label{eq:collectiveH}
\end{align}
Here $G^{rs}$ is the inverse metric, defined on a region $\mathcal Q_m$ where $G$ is positive definite. The classical trajectories are geodesics of this metric~\cite{arnold}.

The quantum Hilbert space and Hamiltonian are
\begin{equation}
 \mathcal H=L^2(\mathcal Q_m,\dd\mu_G),\qquad
 \dd\mu_G=\sqrt{\det G}\,\dd^mq,
 \label{eq:hilbert}
\end{equation}
\begin{equation}
 \widehat H=-\frac{\hbar^2}{2}\Delta_G,
 \qquad
 \Delta_G=\frac1{\sqrt{\det G}}\partial_r
       \left(\sqrt{\det G}\,G^{rs}\partial_s\right).
 \label{eq:laplaceH}
\end{equation}
We use the Friedrichs realization associated with the nonnegative kinetic form
\begin{equation}
 \mathfrak h[\Psi]=\frac{\hbar^2}{2}\int\dd\mu_G\,
 G^{rs}(\partial_r\Psi)^*\partial_s\Psi,
 \label{eq:kineticform}
\end{equation}
with the boundary conditions specified for each model.

The observed modes define a subsystem $S$, and the remaining modes form its environment $E$. On a product coordinate domain, the unitary map $W\Psi=(\det G)^{1/4}\Psi$ gives the factorization $\mathcal H_S\otimes\mathcal H_E$ in the flat-measure representation. For an environment state $\sigma_E$ independent of the input $\rho_S$, the reduced dynamics is
\begin{equation}
 \mathcal E_t(\rho_S)=\Tr_E
 \left[U(t)(\rho_S\otimes\sigma_E)U(t)^\dagger\right],
 \qquad U(t)=e^{-it\widehat H_{\rm flat}/\hbar},
 \label{eq:channel}
\end{equation}
where $\widehat H_{\rm flat}=W\widehat H W^{-1}$. The accessible observables specify the subsystem decomposition~\cite{zanardi}. In our examples, passive material markers fix the observed coordinate and its phase reference. The defining functionals and observable algebras are given in Appendix~\ref{app:preparations}.

\subsection{Cyclic coordinates and quantum coherence}
\label{sec:connection}
Let $q\in\mathbb R$ denote the observed coordinate and $y^a$ the unobserved coordinates. If $G$ is independent of $q$, the coordinate is cyclic and its conjugate momentum $p$ is conserved. Writing $p=\pm k$, with $k>0$, gives
$L^2(\mathbb R,\dd p)\simeq\mathbb C^2_{\rm sign}\otimes L^2(\mathbb R_+,\dd k)$. We encode a logical qubit in the sign degree of freedom, with the joint state of $k$ and the environment prepared independently of the logical input. The logical state at time $t$ is obtained by tracing out these degrees of freedom.

The metric can be written as
\begin{align}
 A_a&=\frac{G_{qa}}{G_{qq}},&
 h_{ab}&=G_{ab}-\frac{G_{qa}G_{qb}}{G_{qq}},\notag\\
 \dd s^2&=G_{qq}(\dd q+A)^2+h_{ab}\dd y^a\dd y^b,
 &A&=A_a\dd y^a,&F&=\dd A.
 \label{eq:connectionmetric}
\end{align}
The one-form $\dd q+A$ is the mechanical connection for translations of $q$, with local connection form $A$ in the chosen coordinates. Its horizontal directions are orthogonal to $\partial_q$ in the kinetic metric~\cite{marsdenphases,littlejohn}. With $h^{ab}$ the inverse of $h_{ab}$,
\begin{equation}
 G^{ab}=h^{ab},\qquad G^{qa}=-h^{ab}A_b,\qquad
 G^{qq}=\frac1{G_{qq}}+A_ah^{ab}A_b,\qquad
 \mu=\sqrt{G_{qq}\det h}.
 \label{eq:connectioninverse}
\end{equation}

For a given observed momentum $p$, the unobserved modes evolve under a conditional Hamiltonian. We impose Dirichlet boundary conditions on a bounded connected Lipschitz domain $\Omega$ and assume a smooth positive definite metric on a neighborhood of $\overline\Omega$. A real bounded potential $V_0(y)$ may be included and is set to zero in the examples. Substituting $\partial_q=ip/\hbar$ gives
\begin{equation}
 H_\mu(p)=-\frac{\hbar^2}{2\mu}
 D_a^{(p)}\!\left(\mu h^{ab}D_b^{(p)}\right)
       +\frac{p^2}{2G_{qq}}+V_0,
 \qquad D_a^{(p)}=\partial_a-\frac{ip}{\hbar}A_a.
 \label{eq:magneticconditional}
\end{equation}
The unitary map $\psi\mapsto\mu^{1/2}\psi$ changes the measure on the unobserved coordinates from $\mu\,\dd y$ to $\dd y$. In this representation,
\begin{align}
 H(p)&=H_E+\frac{p^2}{2}G^{qq}+pB,\notag\\
 H_E&=-\frac{\hbar^2}{2}(\partial_a+l_a)h^{ab}(\partial_b-l_b)+V_0,
 &l_a&=\tfrac12\partial_a\log\mu,\notag\\
 B&=\tfrac12\{c^a,\pi_a\}
   =-i\hbar\left(c^a\partial_a+\tfrac12\partial_ac^a\right),
 &c^a&=-h^{ab}A_b,\quad\pi_a=-i\hbar\partial_a.
 \label{eq:generalB}
\end{align}
Repeated indices $a,b$ are summed, and $\{X,Y\}=XY+YX$. The kinetic quadratic form fixes the symmetric ordering of $B$ and the common form domain of $H(p)$ and $H(-p)$. The term $pB$ is the only contribution that changes sign under $p\mapsto-p$.

As a differential operator, $B$ vanishes if and only if its first-order coefficients $-i\hbar c^a$ vanish identically. Since $c^a=-h^{ab}A_b$ and $h_{ab}$ is positive definite, this is equivalent to $A=0$. Together with $H(p)-H(-p)=2pB$, this yields
\begin{equation}
 A=0\quad\Longleftrightarrow\quad B=0
 \quad\Longleftrightarrow\quad H(p)=H(-p)\quad(p\ne0).
 \label{eq:signcriterion}
\end{equation}
Under the conditions in Eq.~\eqref{eq:signcriterion}, the Hamiltonian acts trivially on the sign degree of freedom. With the observed coordinate and logical observables fixed, this is an exact noiseless subsystem for any joint state of $k$ and the environment independent of the logical input.

A redefinition $q'=q+\Lambda(y)$ acts on each momentum fiber as
\begin{equation}
 \psi'_p(y)=e^{-ip\Lambda(y)/\hbar}\psi_p(y),\qquad
 A'=A-\dd\Lambda.
 \label{eq:markergauge}
\end{equation}
When $A=\dd\Lambda$, the local form satisfies $A'=0$ and the mixed metric components vanish in the new coordinates. This requires $F=\dd A=0$, which is sufficient on a contractible domain and must be supplemented by vanishing periods of $A$ on a multiply connected domain. Since the transformation mixes observed and unobserved coordinates, the same experiment requires transforming both the initial state and the measured observables. Preparing a new product state in the transformed coordinates changes the physical preparation.

Let $\Sigma$ denote the initial joint state of the momentum magnitude and environment, and $H_\pm=H(\pm\widehat k)$, where $\widehat k$ multiplies by $k$. The coherence factor and short-time decoherence coefficient $C_t$ are
\begin{equation}
 f_\Sigma(t)=\Tr\!\left(e^{-itH_+/\hbar}\Sigma e^{itH_-/\hbar}\right),\qquad
 |f_\Sigma(t)|^2=1-C_t t^2+o(t^2),\qquad
 C_t=\frac4{\hbar^2}\Var_\Sigma(\widehat kB).
 \label{eq:generalcurvatureloss}
\end{equation}
The expansion requires finite second moments of $H_+$ and $H_-$. For $\Sigma=\ket g\bra g\otimes\sigma_E$, where $g(k)$ is a normalized wavefunction of the momentum magnitude, we write $f_g=f_\Sigma$. Appendix~\ref{app:geometry} derives the expansion and constructs smooth finite-energy states with $C_t>0$ whenever $B\ne0$. With the observed coordinate and logical observables fixed, $A=0$ is therefore necessary and sufficient for exact preservation of sign coherence at all times for every admissible $\Sigma$ prepared independently of the logical input.

\subsection{Geometric origin of the classical and quantum couplings}
\label{sec:sharedgeometry}
The derivatives of the kinetic metric govern both the acceleration in configuration space and the curvature of the mechanical connection. For a configuration map $\mathbf X$, define
\begin{equation}
 \mathbf e_A=\partial_A\mathbf X,\qquad
 \mathbf e_{BC}=\partial_B\partial_C\mathbf X,\qquad
 \mathcal T_{A|BC}=\rho_0\int_M
           \mathbf e_A\cdot\mathbf e_{BC}\,\dd^3a.
 \label{eq:kineticintegrals}
\end{equation}
Differentiating the metric yields
\begin{align}
 \partial_CG_{AB}&=\mathcal T_{A|BC}+\mathcal T_{B|AC},\notag\\
 \Gamma_{A,BC}
 &=\tfrac12(\partial_BG_{AC}+\partial_CG_{AB}-\partial_AG_{BC})
   =\mathcal T_{A|BC},\notag\\
 G_{AB}\ddot q^B&=-\mathcal T_{A|BC}\dot q^B\dot q^C.
 \label{eq:sharedchristoffel}
\end{align}
The derivatives of the kinetic metric also enter the divergence term in $B$ and the connection curvature $F$,
\begin{align}
 \partial_CG^{qa}&=-G^{qA}(\partial_CG_{AB})G^{Ba},\notag\\
 F_{ab}&=\frac{\mathcal T_{b|qa}-\mathcal T_{a|qb}}{G_{qq}}
       -\frac{G_{qb}\partial_aG_{qq}-G_{qa}\partial_bG_{qq}}{G_{qq}^2}.
 \label{eq:sharedcurvature}
\end{align}
At a point where $G_{qa}=0$, $F_{ab}$ reduces to the antisymmetric combination $(\mathcal T_{b|qa}-\mathcal T_{a|qb})/G_{qq}$.

\section{Symmetry and coherence in the two-shear model}
\label{sec:two}
\subsection{Hamiltonian}
On a periodic domain of side $2\pi\ell$, use dimensionless material coordinates $(x,y,z)=\mathbf a/\ell$ and choose
\begin{align}
 \phi_1=-\cos y,&\quad\chi_1=\sin z,
 &\mathbf w_1&=(\sin y\cos z,0,0),\notag\\
 \phi_2=-\cos x,&\quad\chi_2=\sin y,
 &\mathbf w_2&=(0,0,\sin x\cos y).
 \label{eq:shears}
\end{align}
The exact flows are $\Phi_1^q(x,y,z)=(x+q\sin y\cos z,y,z)$ and $\Phi_2^r(x,y,z)=(x,y,z+r\sin x\cos y)$. With $f=\sin y\cos z$ and $u=x+qf$, their composition is
\begin{equation}
 \mathbf X_{q,r}(\ell x,\ell y,\ell z)=\ell(u,y,z+r\sin u\cos y).
 \label{eq:two_map}
\end{equation}
The kinetic metric obtained from the tangent vectors in Appendix~\ref{app:geometry} is
\begin{equation}
 G=\mathcal I\begin{pmatrix}a(r)&0\\0&1\end{pmatrix},\qquad
 a(r)=1+\frac{r^2}{8},\qquad
 \mathcal I=\frac{\rho_0(2\pi\ell)^3\ell^2}{4}.
 \label{eq:two_metric}
\end{equation}
The classical Hamiltonian $H=p_q^2/(2\mathcal I a(r))+p_r^2/(2\mathcal I)$ couples the modes through the $r$-dependent inertia $\mathcal I a(r)$. Using the Friedrichs realization of the kinetic form, the quantum model is
\begin{align}
 \mathcal H&=L^2(\mathbb R,\dd q)\otimes
        L^2(\mathbb R,\sqrt{a(r)}\dd r)
        =\mathcal H_q\otimes\mathcal H_r,\notag\\
 \widehat H&=\widehat K_q\otimes a(\widehat r)^{-1}
                  +\mathbf1\otimes\widehat K_r,
 \label{eq:two_H}\\
 \widehat K_q&=-\frac{\hbar^2}{2\mathcal I}\partial_q^2,\qquad
 \widehat K_r=-\frac{\hbar^2}{2\mathcal I\sqrt{a(r)}}
                         \partial_r\bigl(\sqrt{a(r)}\partial_r\bigr).
 \notag
\end{align}

\subsection{Coherence between opposite momenta}
\label{sec:kernel}
The Hamiltonian depends on the observed momentum only through $p_q^2$. In the momentum representation,
\begin{equation}
 L^2(\mathbb R,\dd p_q)\simeq\mathbb C^2_{\rm sign}\otimes L^2(\mathbb R_+,\dd k),
 \qquad\widetilde\psi\mapsto
 \begin{pmatrix}\widetilde\psi(k)\\\widetilde\psi(-k)\end{pmatrix}.
 \label{eq:signfactor}
\end{equation}
\begin{align}
\widehat H&=\mathbf1_{\rm sign}\otimes H_{kr},
&
H_{kr}&=\frac{\widehat k^2}{2\mathcal I}a(\widehat r)^{-1}
          +\widehat K_r,\notag\\
\mathcal E_t&=\id_{\rm sign}\otimes\mathcal F_t,
&
\mathcal F_t(\rho)&=\Tr_r\left[
e^{-itH_{kr}/\hbar}(\rho\otimes\sigma_r)
e^{itH_{kr}/\hbar}\right].
\label{eq:noiseless}
\end{align}
Here $H_{kr}$ acts on $L^2(\mathbb R_+,\dd k)\otimes\mathcal H_r$, $\rho$ is a density operator for the momentum magnitude, and $\sigma_r$ is the initial state of the unobserved shear. The sign of momentum defines an exact noiseless subsystem, whose correlations with an external reference are also preserved~\cite{lidar}.

A normalized wavepacket $g\in C_c^\infty(\mathbb R_+)$ of the momentum magnitude, with $C_c^\infty$ denoting smooth compactly supported functions, encodes a logical state $c_+\ket0+c_-\ket1$ as
\begin{equation}
 \widetilde\psi(p)=
 \begin{cases}c_+g(p),&p>0,\\c_-g(-p),&p<0.\end{cases}
 \label{eq:signcode}
\end{equation}
with $|c_+|^2+|c_-|^2=1$. This wavepacket has finite energy when the unobserved mode has finite mean kinetic energy. Decoding traces out the momentum magnitude and the $r$ mode. Relative to the origin fixed by the material markers, the logical Pauli algebra is generated by
\begin{equation}
 Z_{\mathrm L}=\operatorname{sgn}(\widehat p_q),\qquad
 (X_{\mathrm L}\psi)(q)=\psi(-q).
 \label{eq:logicalobs}
\end{equation}

\begin{figure}[!ht]
 \centering
 \includegraphics[width=.53\linewidth]{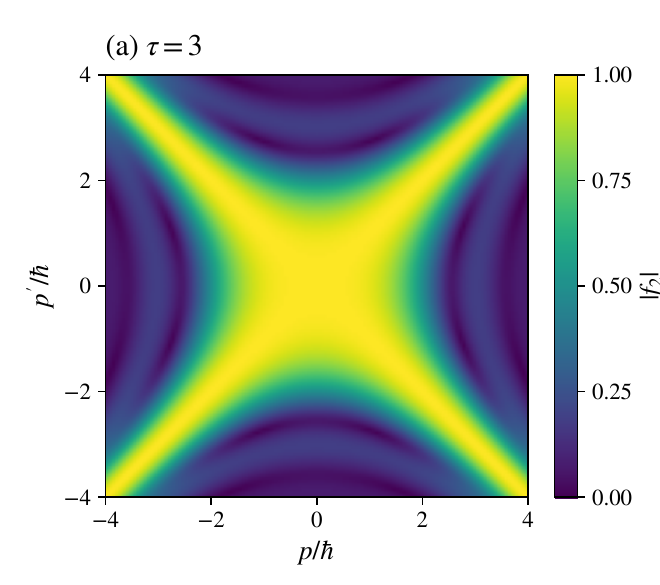}\hfill
 \includegraphics[width=.42\linewidth]{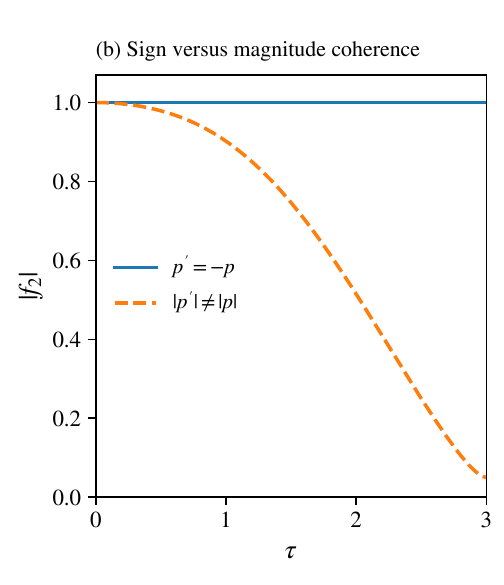}
 \caption{Momentum coherence in the two-shear model. (a) $|f_2(p,p';\tau)|$ at $\tau=3$. The kernel equals unity on the diagonal and anti-diagonal, reflecting conservation of momentum populations and coherence between opposite momenta. (b) Coherence between opposite momenta and between the momentum states $p=\pi\hbar/6$, $p'=5\pi\hbar/6$ with different absolute values. Both panels use the Gaussian initial state in Eq.~\eqref{eq:gaussianpreparation}.}
 \label{fig:kernel}
\end{figure}

For an initial pure environment state $\psi_E$, the coherence kernel is
\begin{align}
 H_2(p)&=\widehat K_r+\frac{p^2}{2\mathcal I}a(\widehat r)^{-1},
 &\psi_{E,p}(t)&=e^{-itH_2(p)/\hbar}\psi_E,\notag\\
 f_2(p,p';t)&=\langle\psi_{E,p'}(t)|\psi_{E,p}(t)\rangle,
 &\widetilde\rho_q(p,p';t)&=f_2(p,p';t)\widetilde\rho_q(p,p';0).
 \label{eq:momentumkernel}
\end{align}
Since $H_2(p)=H_2(-p)$, the kernel satisfies $f_2(p,p;t)=f_2(p,-p;t)=1$, preserving populations and coherence between opposite momenta. Coherences between momenta with different absolute values can decay. Figure~\ref{fig:kernel} shows the kernel and its time dependence. We use the dimensionless time
\begin{equation}
 \tau=\frac{t}{t_{\mathrm Q}},\qquad
 t_{\mathrm Q}=\frac{\mathcal I}{\hbar}.
 \label{eq:timescale}
\end{equation}
The unobserved mode is initially in the Gaussian state specified in Appendix~\ref{app:numerics}.

\FloatBarrier

\section{Euler dynamics and quantum coherence}
\label{sec:four}
\subsection{Cancellation of the initial Euler residual}
We first extend the two-shear model by introducing two additional generating flows with independent parameters. The residual is
\begin{equation}
 \mathcal R_{\mathrm E}=\mathbb P\left(\partial_t\mathbf v+
                    (\mathbf v\cdot\nabla)\mathbf v\right),\qquad
 \mathbb P=\mathbf1-\nabla\Delta^{-1}\nabla\cdot,
 \label{eq:residual}
\end{equation}
where $\mathbb P$ is the Leray projection onto divergence-free fields and $\Delta^{-1}$ acts on zero-mean functions on the periodic domain. For $U=0$, the Euler equation considered in Nambu's note requires $\mathcal R_{\mathrm E}=0$~\cite[pp.~5--6]{note}.

Set $\ell=1$ temporarily and take $q=r=0$, $\dot q=\alpha$, $\dot r=\beta$ at $t=0$. The two-shear variational equations imply $\ddot q=\ddot r=0$ at this time, and the material acceleration is
\begin{equation}
 \mathbf A_2=\left.\partial_t^2\mathbf X\right|_0
       =\alpha\beta(0,0,\sin(2y)\cos x\cos z).
 \label{eq:A2}
\end{equation}
With $\psi=\sin(2y)\cos x\sin z/6$, the divergence-free projection yields
\begin{align}
 \mathcal R_{\mathrm E}(0)&=\alpha\beta\mathbf W,\notag\\
 \mathbf W&=\left(\frac{\sin x\sin(2y)\sin z}{6},
       -\frac{\cos x\cos(2y)\sin z}{3},
        \frac{5\cos x\sin(2y)\cos z}{6}\right).
 \label{eq:W}
\end{align}
To include the residual in the tangent space, choose the generating pairs
\begin{align}
 \phi_3&=\tfrac13\cos(2y)\cos z,&\chi_3&=\sin x,\notag\\
 \phi_4&=\tfrac16\sin x\cos z,&\chi_4&=-\tfrac12\cos(2y).
 \label{eq:addedpairs}
\end{align}
The generated fields satisfy $\mathbf W=\mathbf w_3+\mathbf w_4$, with components given in Appendix~\ref{app:geometry}. Their flows define the four-coordinate family
\begin{equation}
 \mathbf X_{q,r,s_3,s_4}
 =\ell\,\Phi_4^{s_4}\circ\Phi_3^{s_3}\circ\Phi_2^r\circ\Phi_1^q,
 \label{eq:fourmap}
\end{equation}
whose metric at the identity is
\begin{equation}
 G(0)=\mathcal I\begin{pmatrix}
 1&0&0&0\\0&1&0&0\\0&0&5/18&1/18\\0&0&1/18&1/36
 \end{pmatrix}.
 \label{eq:G0}
\end{equation}
We take $s_3(0)=s_4(0)=\dot s_3(0)=\dot s_4(0)=0$, keeping the initial values of $q,r,\dot q,\dot r$ specified above. The variational equations for the additional coordinates involve the lower block $G_{\{3,4\}}(0)$,
\begin{equation}
 G_{\{3,4\}}(0)
 \begin{pmatrix}\ddot s_3\\\ddot s_4\end{pmatrix}
 =-\mathcal I\alpha\beta\begin{pmatrix}1/3\\1/12\end{pmatrix}.
 \label{eq:Gramload}
\end{equation}
At the identity, the additional tangent vectors are orthogonal to the two shear tangent vectors. The variational equations give $\ddot q=\ddot r=0$ and $\ddot s_3=\ddot s_4=-\alpha\beta$, cancelling the divergence-free residual,
\begin{equation}
 \mathbf A_4=\mathbf A_2-\alpha\beta\mathbf W
            =\alpha\beta\nabla\psi,\qquad
 \mathcal R_{\mathrm E}^{(4)}(0)=0.
 \label{eq:correctedacc}
\end{equation}
Here $\mathcal R_{\mathrm E}^{(4)}$ is the Euler residual of the four-coordinate trajectory. The remaining gradient acceleration is generated by the pressure $P=-\rho_0\ell^2\alpha\beta\psi$. Let $\mathbf X_2$ and $\mathbf X_4$ denote the two-shear and four-coordinate trajectories, and $\mathbf X_{\mathrm E}$ the Euler map with the same initial data. The short-time position errors are $\mathbf X_2-\mathbf X_{\mathrm E}=\tfrac12t^2\mathcal R_{\mathrm E}(0)+O(t^3)$ and $\mathbf X_4-\mathbf X_{\mathrm E}=O(t^3)$. Appendix~\ref{app:geometry} gives a finite-time velocity bound in terms of the accumulated residual.

\subsection{Connection curvature and momentum reversal}
\label{sec:signcoupling}
The additional coordinates also induce nonzero curvature of the mechanical connection. For $y=(r,s_3,s_4)$, the derivatives of the configuration map at the identity are
\begin{equation}
 \mathbf e_A=\ell\mathbf w_A,\qquad
 \mathbf e_{BC}=\ell D\mathbf w_{\max(B,C)}\mathbf w_{\min(B,C)}.
 \label{eq:sharedflowderivatives}
\end{equation}
Substitution in Eq.~\eqref{eq:kineticintegrals} gives
\begin{equation}
 A=-\frac{s_3}{6}\dd r+\frac r6\dd s_3+
                  \frac r{12}\dd s_4+O(|y|^2),\qquad
 F_0=\frac13\dd r\wedge\dd s_3+\frac1{12}\dd r\wedge\dd s_4.
 \label{eq:sharedA}
\end{equation}
$F_0=F|_{y=0}$ denotes the curvature at the identity. The coefficients $\mathcal T_{A|BC}$ enter both the acceleration term in Eq.~\eqref{eq:Gramload} and the curvature in Eq.~\eqref{eq:sharedcurvature}.

Shifts of $q$ amount to volume-preserving relabelings, so the periodic kinetic metric is independent of $q$. We work on $\mathbb R_q\times\Omega$, where $\Omega$ is a bounded Lipschitz domain containing $0$ and $G$ is smooth and positive definite on a neighborhood of $\overline\Omega$. We impose Dirichlet conditions in the unobserved coordinates. The material markers described in Appendix~\ref{app:preparations} fix the observed coordinate and phase reference.

The conditional Hamiltonian is Eq.~\eqref{eq:generalB} with $V_0=0$. Its leading sign-dependent coupling is
\begin{equation}
 B_1=\frac1{\mathcal I}\left(\frac{s_3\pi_r}{6}-3r\pi_{s_4}\right).
 \label{eq:Blinear}
\end{equation}
The conditional Hamiltonians satisfy $H(p)-H(-p)=2pB$ and therefore differ for opposite nonzero momenta. Nonzero $F_0$ precludes local elimination of $A$ by redefining the cyclic coordinate.

We retain the encoding in Eq.~\eqref{eq:signcode} and prepare $\Sigma=\ket g\bra g\otimes\sigma_E$. The variance in Eq.~\eqref{eq:generalcurvatureloss} becomes
\begin{equation}
 \Var_{\Sigma}(\widehat kB)=\langle k^2\rangle_g\Var_{\sigma_E}(B)
                       +\Var_g(k)\langle B\rangle_{\sigma_E}^2.
 \label{eq:productvariance}
\end{equation}
The moments of $k$ are given by $\langle k^n\rangle_g=\int_0^\infty k^n|g(k)|^2\dd k$ for $n=1,2$, with $\Var_g(k)=\langle k^2\rangle_g-\langle k\rangle_g^2$. For a compactly supported initial state, enlarging the domain leaves $C_t$ unchanged provided the initial state, the metric on its support, and the observed coordinate and phase reference remain fixed. Appendix~\ref{app:preparations} specifies the wavepackets used below.

\subsection{Three-coordinate reduction}
\label{sec:linked}
We restrict the four-coordinate family to the submanifold defined by $s_3=a_\vartheta s$ and $s_4=b_\vartheta s$. For $0\le\vartheta\le\pi/2$, choose
\begin{equation}
 \begin{gathered}
 D_\vartheta=10\cos^2\vartheta+4\sin\vartheta\cos\vartheta+\sin^2\vartheta,\\
 a_\vartheta=\frac{6\cos\vartheta}{\sqrt{D_\vartheta}},\qquad
 b_\vartheta=\frac{6\sin\vartheta}{\sqrt{D_\vartheta}}.
 \end{gathered}
 \label{eq:directionnormalization}
\end{equation}
The restricted configuration map is
\begin{equation}
 \mathbf X^{(\vartheta)}_{q,r,s}
 =\ell\,\Phi_4^{b_\vartheta s}\circ\Phi_3^{a_\vartheta s}
          \circ\Phi_2^r\circ\Phi_1^q.
 \label{eq:linkedmap}
\end{equation}
The observed coordinate is $q$, and $r,s$ are unobserved. Let $T_\vartheta$ denote the Jacobian of the embedding $(q,r,s)\mapsto(q,r,a_\vartheta s,b_\vartheta s)$. The induced metric is
\begin{equation}
 \begin{gathered}
 G_\vartheta(q,r,s)=T_\vartheta^{\mathsf T}
      G(q,r,a_\vartheta s,b_\vartheta s)T_\vartheta,\\
 G_\vartheta(0)=\mathcal I\,\mathbf1_3.
 \end{gathered}
 \label{eq:linkedmetric}
\end{equation}
We quantize the restricted model using $G_\vartheta$.

Take $q(0)=r(0)=s(0)=\dot s(0)=0$, $\dot q(0)=\alpha$, and $\dot r(0)=\beta$. The initial residual is the component of $\mathcal R_{\mathrm E}^{(2)}(0)$ orthogonal in $L^2$ to $a_\vartheta\mathbf w_3+b_\vartheta\mathbf w_4$. For $\alpha\beta\ne0$, the squared relative residual is
\begin{equation}
 \frac{\|\mathcal R_\vartheta(0)\|_2^2}
      {\|\mathcal R_{\mathrm E}^{(2)}(0)\|_2^2}
 =\frac{2(\cos\vartheta-\sin\vartheta)^2}{5D_\vartheta},
 \label{eq:linkedresidual}
\end{equation}
Here $\mathcal R_\vartheta$ and $\mathcal R_{\mathrm E}^{(2)}$ are the Euler residuals of the reduced and two-shear trajectories, respectively, as defined by Eq.~\eqref{eq:residual}. At $\vartheta=\pi/4$, the additional coordinate cancels the initial Euler residual. Appendix~\ref{app:geometry} gives the corresponding acceleration.

$B_\vartheta$ is obtained from Eq.~\eqref{eq:generalB} using $G_\vartheta^{-1}$. Define
\begin{equation}
 \alpha_\vartheta=\frac{a_\vartheta}{6},\qquad
 \beta_\vartheta=\frac{a_\vartheta}{6}+\frac{b_\vartheta}{12}.
 \label{eq:adaptedcoefficients}
\end{equation}
The leading term of $B_\vartheta$ near the identity is
\begin{equation}
 B_{\vartheta,1}=\frac1{\mathcal I}
       (\alpha_\vartheta s\pi_r-\beta_\vartheta r\pi_s).
 \label{eq:linkedB}
\end{equation}
\subsection{Suppression of short-time decoherence}
\label{sec:adapted}
The differential expression for $B_{\vartheta,1}$ is invariant under the dilation $r=\delta u$, $s=\delta v$, with $\pi_r=\delta^{-1}\pi_u$ and $\pi_s=\delta^{-1}\pi_v$. For the family constructed below, the kinetic energy associated with localization scales as $\delta^{-2}$ as $\delta\to0$ at fixed angle and profile. For $0\le\vartheta<\pi/2$, the vector field $\alpha_\vartheta s\partial_r-\beta_\vartheta r\partial_s$ has elliptical integral curves given by the level sets of the quadratic form
\begin{equation}
 Q_\vartheta=\beta_\vartheta r^2+\alpha_\vartheta s^2,
 \qquad
 (\alpha_\vartheta s\partial_r-\beta_\vartheta r\partial_s)Q_\vartheta=0.
 \label{eq:adaptedinvariant}
\end{equation}
An amplitude constant along these ellipses is annihilated by $B_{\vartheta,1}$. We call such amplitudes adapted.

For a smooth real profile $\eta(\xi)$ that vanishes for $\xi\ge1$, take
\begin{equation}
 \psi^0_{\delta,\vartheta}(r,s)=\mathcal N
       \eta(Q_\vartheta/\delta^2),\qquad
 \|\psi^0_{\delta,\vartheta}\|_2=1,
 \label{eq:adaptedstate}
\end{equation}
with support inside the environment domain. As shown in Appendix~\ref{app:geometry}, for $\delta\to0$ at fixed $\vartheta$ and $\eta$,
\begin{equation}
 B_{\vartheta,1}\psi^0_{\delta,\vartheta}=0,\qquad
 \|B_\vartheta\psi^0_{\delta,\vartheta}\|_2
       =O(\hbar\delta/\mathcal I).
 \label{eq:adaptedbound}
\end{equation}
For $\delta$-independent $p_{r0}$, the wavefunction $e^{ip_{r0}r/\hbar}\psi^0_{\delta,\vartheta}$ has mean momentum $p_{r0}$ in the $r$ direction and satisfies
\begin{equation}
 B_{\vartheta,1}\bigl(e^{ip_{r0}r/\hbar}\psi^0_{\delta,\vartheta}\bigr)
 =\frac{\alpha_\vartheta p_{r0}s}{\mathcal I}
       e^{ip_{r0}r/\hbar}\psi^0_{\delta,\vartheta}.
 \label{eq:adaptedphase}
\end{equation}
For this wavefunction, Eqs.~\eqref{eq:adaptedbound} and \eqref{eq:adaptedphase} imply $C_t=O(\delta^2)$ with the distribution of $k$ and mean momentum held fixed. For the restricted model, the connection curvature at $r=s=0$ is $(\alpha_\vartheta+\beta_\vartheta)\dd r\wedge\dd s$ and remains nonzero. Section~\ref{sec:quantumnumeric} compares these states at equal mean energy.

\section{Classical accuracy and quantum coherence}
\label{sec:numerical}
We use the dimensionless time $\tau$ defined in Eq.~\eqref{eq:timescale}, energies in units of $E_{\mathrm Q}=\hbar^2/\mathcal I$, and momenta in units of $\hbar$. The notation $f_g(\tau)$ denotes the coherence factor evaluated at $t=t_{\mathrm Q}\tau$. Its short-time expansion is
\begin{equation}
 1-|f_g(\tau)|^2=C_\tau\tau^2+o(\tau^2),\qquad
 C_\tau=t_{\mathrm Q}^2C_t.
 \label{eq:timeunits}
\end{equation}
\subsection{Comparison with Euler dynamics}
\label{sec:classicalcomparison}
Cancellation of the initial Euler residual eliminates the linear term in the velocity error, yielding the $O(\tau^2)$ bound derived in Appendix~\ref{app:geometry}. Figure~\ref{fig:residual} compares the two-shear, three-coordinate ($\vartheta=\pi/4$), and four-coordinate models. All trajectories start from the identity map with $\dd q/\dd\tau=\dd r/\dd\tau=1$ and zero velocities in the additional coordinates. The model trajectories satisfy Eq.~\eqref{eq:sharedchristoffel} and are compared with an independent numerical solution of the Euler equation. We measure lengths in units of $\ell$ and velocities in units of $\ell/t_{\mathrm Q}$. The velocity error is measured in the root-mean-square norm $\|f\|_{\rm rms}=[(2\pi)^{-3}\int|f|^2\dd^3x]^{1/2}$.

\begin{figure}[!htbp]
 \centering
 \includegraphics[width=.48\linewidth]{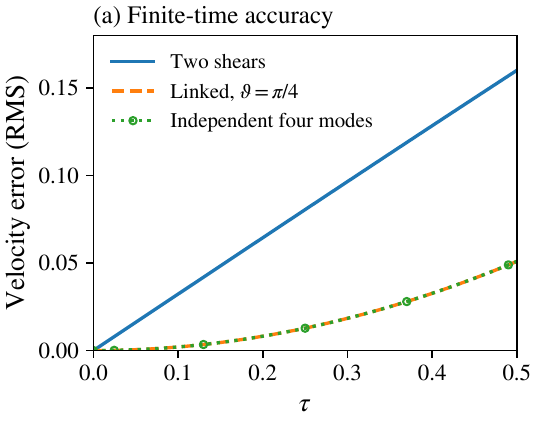}\hfill
 \includegraphics[width=.48\linewidth]{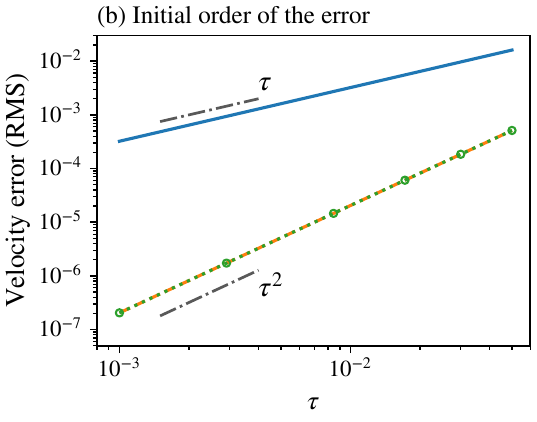}
 \caption{Velocity error relative to the Euler solution for identical initial data. (a) RMS velocity errors for the two-shear, three-coordinate ($\vartheta=\pi/4$), and four-coordinate models. The latter two curves nearly coincide. (b) Numerical short-time scaling, linear for two shears and quadratic for the three- and four-coordinate models. Reference slopes indicate these orders.}
 \label{fig:residual}
\end{figure}

The additional coordinate reduces the velocity error over the interval considered. Allowing the two added flow parameters to vary independently has little effect on the velocity error.

\subsection{Initial states at equal mean energy}
\label{sec:quantumnumeric}
We compare product and adapted states at equal mean energy for $\vartheta=\pi/4$, where the initial Euler residual vanishes. The product state has an environment wavefunction that factorizes in $r$ and $s$. Both states use the wavepacket in $k=|p|$ defined in Eq.~\eqref{eq:wavepackets}, with $k_0=p_{r0}=1$, $\Delta k=0.25$, and $\langle\pi_s\rangle=0$.

The product wavefunction is
\begin{equation}
 \psi_E(r,s)=\delta^{-1}
 w_0(r/\delta)w_0(s/\delta)e^{ip_{r0}r/\hbar},
 \label{eq:linkedpacket}
\end{equation}
where $\delta>0$ is the width and $w_0$ is the normalized profile in Eq.~\eqref{eq:bump}. For the adapted state, we multiply the amplitude in Eq.~\eqref{eq:adaptedstate} by $e^{ip_{r0}r/\hbar}$ and choose the profile
\begin{equation}
 \eta(\xi)=\exp\!\left[-\frac{1}{2(1-\xi)}\right]\quad(0\le\xi<1),\qquad
 \eta(\xi)=0\quad(\xi\ge1).
 \label{eq:adaptedprofile}
\end{equation}

\begin{figure}[!htbp]
 \centering
 \includegraphics[width=.48\linewidth]{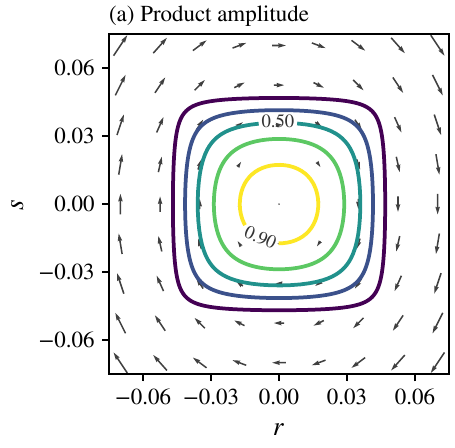}\hfill
 \includegraphics[width=.48\linewidth]{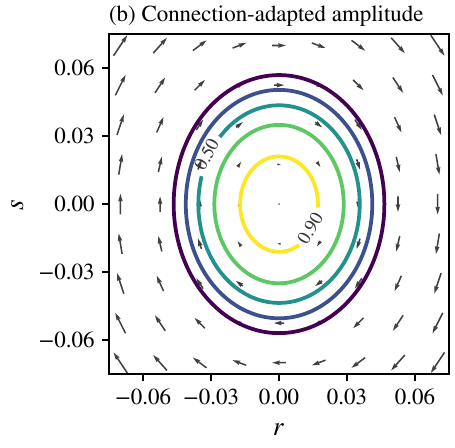}
 \caption{Probability densities of the (a) product and (b) adapted states at $\vartheta=\pi/4$ and $\overline E=10^3\,E_{\mathrm Q}$. Both wavefunctions have the phase factor $e^{ir}$. Each density is divided by its maximum. Arrows indicate $(\alpha_\vartheta s,-\beta_\vartheta r)$, along whose integral curves the adapted amplitude is constant.}
 \label{fig:stateshape}
\end{figure}

For each preparation, the width is chosen to give the prescribed mean energy, evaluated with $G_\vartheta$. Figure~\ref{fig:stateshape} shows the probability densities. Table~\ref{tab:shapedesign} lists $C_\tau$ evaluated with the full operator $B_\vartheta$. The coefficient is more than three orders of magnitude smaller for the adapted state at both energies. For the adapted family, the energy dependence is consistent with the asymptotic bound $C_\tau=O(E_{\mathrm Q}/\overline E)$ derived in Appendix~\ref{app:geometry}.

\begin{table}[!htbp]
 \centering
 \begin{tabular}{ccc}\toprule
 $\overline E/E_{\mathrm Q}$ & Product $C_\tau$ & Adapted $C_\tau$\\\midrule
 $10^3$ & $0.23$ & $\scinum{1.6}{-4}$\\
 $\scinum{7.3}{3}$ & $0.23$ & $\scinum{2.2}{-5}$\\\bottomrule
 \end{tabular}
 \caption{Coefficients $C_\tau$ for product and adapted states at equal mean energy and $\vartheta=\pi/4$. The energies are matched before rounding, and the coefficients are evaluated with the full metric.}
 \label{tab:shapedesign}
\end{table}

\FloatBarrier

\subsection{Finite-time decoherence}
\label{sec:linkedtime}
We evolve the two states in Figure~\ref{fig:stateshape} with the conditional Hamiltonians $H(\pm k)$, using the full kinetic metric and a sine basis satisfying Dirichlet boundary conditions. Averaging their conditional overlaps over $k$ with weight $|g(k)|^2$ yields $f_g$. For the logical density matrix $\rho_{\rm L}$ in the basis of momentum signs, the dephasing channel and coherence loss are
\begin{equation}
 \mathcal D_{f_g}(\rho_{\rm L})=
 \begin{pmatrix}
  \rho_{++}&f_g\rho_{+-}\\ f_g^*\rho_{-+}&\rho_{--}
 \end{pmatrix},\qquad
 \mathcal L(\tau)=1-|f_g(\tau)|^2.
 \label{eq:logicalchannelmatrix}
\end{equation}

\begin{figure}[!htbp]
 \centering
 \includegraphics[width=.49\linewidth]{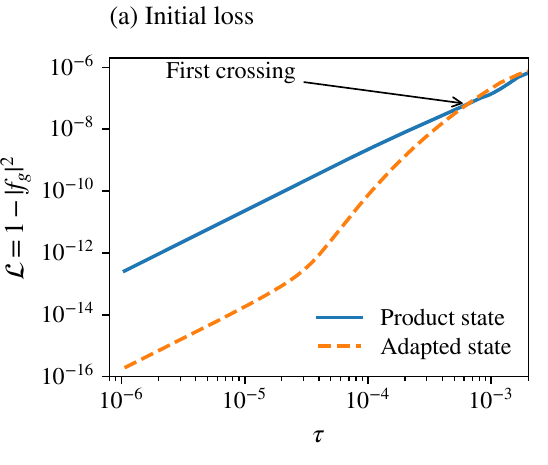}\hfill
 \includegraphics[width=.49\linewidth]{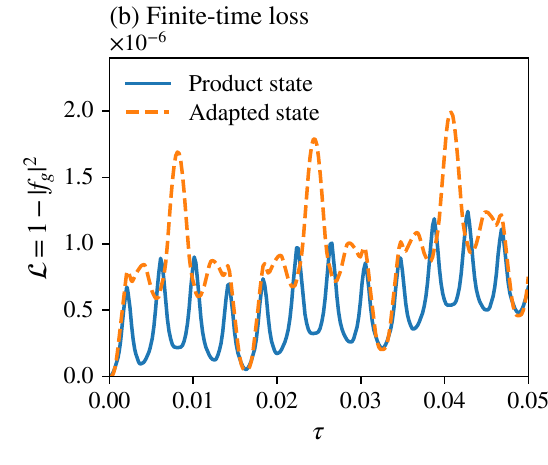}\\[2mm]
 \includegraphics[width=.49\linewidth]{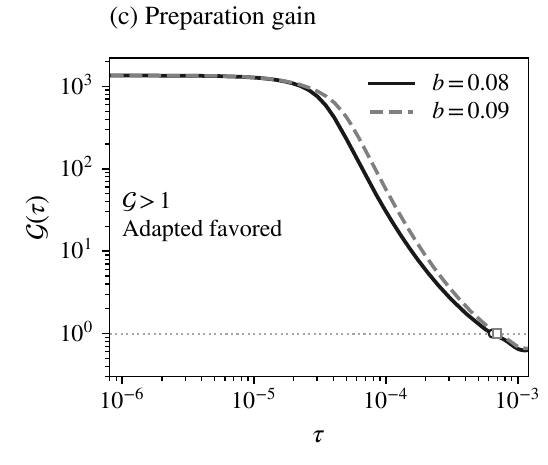}\hfill
 \includegraphics[width=.49\linewidth]{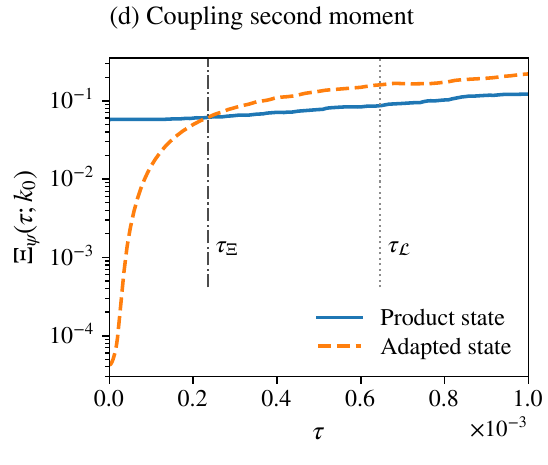}
 \caption{Evolution of the states in Figure~\ref{fig:stateshape}. Blue solid and orange dashed curves denote product and adapted states. (a,b) Short-time and finite-time coherence loss on $\Omega=(-b,b)^2$ with $b=0.08$. (c) Ratio of coherence losses for the same initial states on domains with $b=0.08$ and $0.09$. (d) Dimensionless second moments $\Xi_\psi$ for evolution under $H_{\rm c}(k_0)$ on the domain with $b=0.08$. The vertical lines mark the first crossing of the second moments ($\tau_\Xi$) and of the coherence losses ($\tau_{\mathcal L}$).}
 \label{fig:linkedtime}
\end{figure}

We compare the losses through their ratio,
\begin{equation}
 \mathcal G(\tau)=
 \frac{\mathcal L_{\rm product}(\tau)}{\mathcal L_{\rm adapted}(\tau)}
 \qquad\bigl(\mathcal L_{\rm adapted}(\tau)>0\bigr).
 \label{eq:preparationgain}
\end{equation}
Values $\mathcal G>1$ correspond to a smaller loss for the adapted state. The first crossing in Figure~\ref{fig:linkedtime}(a) corresponds to $\mathcal G=1$ in Figure~\ref{fig:linkedtime}(c). Both states exhibit partial revivals at later times. The adapted state has a larger maximum loss over the sampled times in Figure~\ref{fig:linkedtime}(b).

We decompose the conditional Hamiltonians into their even and odd parts in momentum and define the coupling operator in the interaction picture.
\begin{equation}
 \begin{split}
 H_{\rm c}(k)&=H_E+\frac{k^2}{2}G^{qq}(y),
 \qquad H(\pm k)=H_{\rm c}(k)\pm kB_\vartheta,\\
 B_{I,k}(t)&=e^{itH_{\rm c}(k)/\hbar}B_\vartheta e^{-itH_{\rm c}(k)/\hbar}.
 \end{split}
 \label{eq:fullcommon}
\end{equation}
The conditional overlap follows from the time-ordered evolution generated by $B_{I,k}$, as derived in Appendix~\ref{app:geometry}. Near the identity, the principal kinetic term is $H_{E,0}=(\pi_r^2+\pi_s^2)/(2\mathcal I)$, and
\begin{equation}
 [H_{E,0},B_{\vartheta,1}]
 =\frac{i\hbar(\beta_\vartheta-\alpha_\vartheta)}{\mathcal I^2}\pi_r\pi_s.
 \label{eq:adaptedcommutator}
\end{equation}
At $\vartheta=\pi/4$, $\pi_r\pi_s\psi^0_{\delta,\vartheta}\ne0$ for the profile in Eq.~\eqref{eq:adaptedprofile}. The initial amplitude therefore leaves the kernel of $B_{\vartheta,1}$ at short times under evolution generated by $H_{E,0}$. The subsequent evolution depends on the full Hamiltonian and boundary conditions.

For either initial wavefunction $\psi_E$, we evaluate the second moment of $B_\vartheta$ after evolution under $H_{\rm c}(k)$,
\begin{equation}
 \Xi_\psi(t;k)=\frac{\mathcal I^2}{\hbar^2}
       \left\|B_\vartheta e^{-itH_{\rm c}(k)/\hbar}\psi_E\right\|_2^2.
 \label{eq:commonresponse}
\end{equation}
At $t=0$,
\begin{equation}
 C_\tau=4\frac{\langle k^2\rangle_g}{\hbar^2}\,\Xi_\psi(0;k_0)
 \qquad\text{when }\langle B_\vartheta\rangle_{\psi_E}=0.
 \label{eq:responseinitialrelation}
\end{equation}
The curves for $\Xi_\psi(t;k_0)$ cross before the coherence-loss curves [Figure~\ref{fig:linkedtime}(d)]. The coherence loss follows from the full conditional overlaps averaged over $k$ with weight $|g(k)|^2$.

Increasing the domain half-width from $b=0.08$ to $0.09$ delays the first crossing of the coherence-loss curves without changing the initial states or $C_\tau$ [Figure~\ref{fig:linkedtime}(c)]. The numerical methods are described in Appendix~\ref{app:numerics}, and convergence of the dephasing channel is established in Appendix~\ref{app:channelproof}.
\FloatBarrier

\section{Discussion}
\label{sec:discussion}
The kinetic metric induced by fluid deformations governs their classical dynamics and the coupling between quantized modes. In the shear models studied here, additional coordinates cancel the initial Euler residual, eliminating the linear term in the velocity error. The numerical results show quadratic short-time scaling for the three- and four-coordinate models. The resulting mechanical connection has nonzero curvature, which obstructs local elimination of the mixed metric components by redefining the cyclic coordinate.

For the states considered here, amplitudes in the kernel of the leading-order coupling suppress $C_\tau$ even in the presence of nonzero connection curvature. Coherence at later times depends on the conditional evolution of the unobserved modes. The interaction-picture coupling is generally time dependent because $H_{\rm c}(k)$ and $B_\vartheta$ do not commute, and the conditional overlaps involve time-ordered evolution~\cite{gorin}. In the numerical example, the product state retains greater coherence over some later intervals despite its larger initial coefficient.

Once the observed subsystem is specified, this construction applies to the Laplace--Beltrami quantization of kinetic Hamiltonians with cyclic coordinates and coordinate-dependent inertia. Nuclear collective motion offers a further setting in which to investigate the relation between kinetic geometry and quantum coherence.

\section*{Acknowledgment}
The author thanks Yutaka Hosotani for discussions and for his work preserving Nambu's unpublished notes at the Yoichiro Nambu Memorial, Osaka University.

This work was supported by the U.S. Department of Energy, Office of Science, under Contract No. DE-SC0026415. The quantum information aspects of this study were supported in part by the U.S. National Science Foundation under Grant No. OSI-2328774. 

\FloatBarrier
\appendix
\addtocontents{toc}{\protect\setcounter{tocdepth}{1}}

\section{Supplementary derivations}
\label{app:geometry}
Nambu's note uses material coordinates $(\xi,\eta,\zeta)$ and spatial positions $(x,y,z)$~\cite{note}. The mass density and potential energy density in Eq.~\eqref{eq:action} are defined per unit reference volume. For $F_{iA}=\partial_{a^A}X^i$ and $C_{iA}=(\cof F)_{iA}$, the identities $\delta J=C_{iA}\partial_{a^A}\delta X^i$ and $\partial_{a^A}C_{iA}=0$ give
\begin{equation}
 \rho_0\partial_t^2X^i
 =-\partial_{X^i}U-\partial_{a^A}(\lambda C_{iA}).
 \label{eq:materialEuler}
\end{equation}
On $J_{\mathbf X}=1$, the definition $P=\lambda\circ\mathbf X^{-1}$ gives $\partial_{a^A}(\lambda C_{iA})=\partial_{x^i}P$ and hence Eq.~\eqref{eq:euler}.

The three-dimensional generator relation in the note uses the Jacobian bracket $[f,g,h]_{\mathbf a}=\partial(f,g,h)/\partial(a^1,a^2,a^3)$ in the material coordinates~\cite[p.~4]{note}. For Eulerian generators, set $\widetilde\phi=\phi\circ\mathbf X$ and $\widetilde\chi=\chi\circ\mathbf X$. The chain rule gives
\begin{equation}
 [X^i,\widetilde\phi,\widetilde\chi]_{\mathbf a}
 =J_{\mathbf X}\{x^i,\phi,\chi\}_{\mathbf x},\qquad
 \{f,g,h\}_{\mathbf x}=\nabla f\cdot(\nabla g\times\nabla h).
 \label{eq:pullback}
\end{equation}
On volume-preserving configurations, this yields $\mathbf v=\nabla\phi\times\nabla\chi$.

For the two-shear family, the tangent vectors of Eq.~\eqref{eq:two_map} are
\begin{equation}
 \partial_q\mathbf X=\ell(f,0,rf\cos u\cos y),\qquad
 \partial_r\mathbf X=\ell(0,0,\sin u\cos y).
 \label{eq:two_tangents}
\end{equation}
Their kinetic inner products over the periodic domain give $G_{qr}=0$, $G_{rr}=\mathcal I$, and $G_{qq}=\mathcal I(1+r^2/8)$, as in Eq.~\eqref{eq:two_metric}.

The generator pairs in Eq.~\eqref{eq:addedpairs} define the additional fields
\begin{align}
 \mathbf w_3&=\left(0,-\tfrac13\cos x\cos(2y)\sin z,
                       \tfrac23\cos x\sin(2y)\cos z\right),\notag\\
 \mathbf w_4&=\left(\tfrac16\sin x\sin(2y)\sin z,0,
                       \tfrac16\cos x\sin(2y)\cos z\right),
 \qquad\mathbf W=\mathbf w_3+\mathbf w_4.
 \label{eq:addedflows}
\end{align}
Integration over the dimensionless periodic cell using Eq.~\eqref{eq:W} yields
\[
 \|\mathcal R_{\mathrm E}(0)\|_{L^2([0,2\pi)^3)}^2
 =\frac{5\pi^3}{6}\alpha^2\beta^2.
\]
Equations~\eqref{eq:directionnormalization} and \eqref{eq:linkedresidual} imply that the initial Euler residual vanishes at $\vartheta=\pi/4$. At this angle, $a_\vartheta=b_\vartheta=6/\sqrt{15}$ and $\ddot s(0)=-\sqrt{15}\alpha\beta/6$, so the added acceleration is $-\alpha\beta(\mathbf w_3+\mathbf w_4)$.

Let $\mathbf v_m$ be a smooth incompressible velocity field and $\mathbf u$ the Euler solution with the same initial velocity on a common existence interval. Write $\mathcal R_m=\mathbb P[\partial_t\mathbf v_m+(\mathbf v_m\cdot\nabla)\mathbf v_m]$ and $\gamma_m=\|\nabla\mathbf v_m\|_{L^\infty,\mathrm{op}}$, where $\mathrm{op}$ denotes the induced matrix norm. The standard $L^2$ energy estimate on the periodic domain gives
\begin{equation}
 \|\mathbf v_m(t)-\mathbf u(t)\|_2
 \le\int_0^t\exp\left[\int_s^t\gamma_m(v)\dd v\right]
                       \|\mathcal R_m(s)\|_2\dd s.
 \label{eq:residualbound}
\end{equation}

The coefficient $C_t$ follows from an identity for conditional overlaps. For self-adjoint Hamiltonians $H_a,H_b$, an initial state $\sigma$, and $D=H_a-H_b$,
\begin{equation}
 f(t)=\Tr\left(e^{-itH_a/\hbar}\sigma e^{itH_b/\hbar}\right),\qquad
 |f(t)|^2=1-\frac{t^2}{\hbar^2}\Var_\sigma(D)+o(t^2).
 \label{eq:overlaplemma}
\end{equation}
Here $\Var_\sigma(D)=\Tr(\sigma D^2)-[\Tr(\sigma D)]^2$, with second moments understood as quadratic-form expectations. Assuming $\Tr(\sigma H_a^2)+\Tr(\sigma H_b^2)<\infty$, a first-order expansion of the normalized conditional states gives Eq.~\eqref{eq:overlaplemma}. For a pure state, the assumption is $\psi\in\operatorname{Dom}(H_a)\cap\operatorname{Dom}(H_b)$ and the second moment of $D$ is $\|D\psi\|^2$.

For $B\ne0$, choose a real normalized $\psi_E\in C_c^\infty(\Omega)$ supported where the vector field $c=(c^a)$ is nonzero, with $B\psi_E\ne0$, and normalized $g\in C_c^\infty(\mathbb R_+)$. Integration by parts gives $\langle B\rangle_{\psi_E}=0$ and
\begin{equation}
 \Var_{|g\rangle\langle g|\otimes|\psi_E\rangle\langle\psi_E|}(\widehat kB)
       =\langle k^2\rangle_g\|B\psi_E\|_2^2>0.
 \label{eq:realstateloss}
\end{equation}
For fixed $0\le\vartheta<\pi/2$ and profile $\eta$, the linear part of the vector field $c$ in $B_\vartheta$ is $c_1=(\alpha_\vartheta s,-\beta_\vartheta r)/\mathcal I$, with $y=(r,s)$. Smoothness implies
\[
 \begin{aligned}
 c-c_1&=O(|y|^2/\mathcal I),\\
 \nabla\cdot(c-c_1)&=O(|y|/\mathcal I).
 \end{aligned}
\]
The amplitude in Eq.~\eqref{eq:adaptedstate} has support $|y|=O(\delta)$ and gradient norm $\|\nabla\psi^0_{\delta,\vartheta}\|_2=O(\delta^{-1})$. The differential and multiplication terms in $(B_\vartheta-B_{\vartheta,1})\psi^0_{\delta,\vartheta}$ are therefore both $O(\hbar\delta/\mathcal I)$ in $L^2$, proving Eq.~\eqref{eq:adaptedbound}. At fixed $p_{r0}$ and $g$, the phase in Eq.~\eqref{eq:adaptedphase} contributes $O(|p_{r0}|\delta/\mathcal I)$, so $C_t=O(\delta^2)$.

Since $G_\vartheta(0)=\mathcal I\mathbf1_3$, Eq.~\eqref{eq:initialenergy} implies $\overline E\sim\hbar^2\|\nabla\psi^0_{\delta,\vartheta}\|_2^2/(2\mathcal I)\propto\delta^{-2}$. Consequently, $C_\tau=O(E_{\mathrm Q}/\overline E)$ for this family of states as $\delta\to0$, with the geometry, profile, $g$, and $p_{r0}$ fixed.

Using the Hamiltonian $H_{\rm c}(k)$ in Eq.~\eqref{eq:fullcommon}, factor the conditional propagators $U_\pm(t;k)=e^{-itH(\pm k)/\hbar}$ as $U_\pm(t;k)=e^{-itH_{\rm c}(k)/\hbar}V_\pm(t;k)$. Then
\begin{equation}
 V_\pm(t;k)=\mathcal T\exp\left[\mp\frac{ik}{\hbar}
                       \int_0^t B_{I,k}(s)\dd s\right],
 \label{eq:interactionhistory}
\end{equation}
where $\mathcal T$ denotes time ordering. The unitary factors generated by $H_{\rm c}(k)$ cancel in the overlap, whose evolution is governed by $B_{I,k}$.

\section{Observable coordinates and state preparations}
\label{app:preparations}
In the original weighted Hilbert space, the subsystem decomposition in Section~\ref{sec:reduced} is specified by the observable algebras~\cite{zanardi}
\begin{equation}
 \mathfrak A_S=W^{-1}\bigl(\mathcal B(\mathcal H_S)\otimes\mathbf1\bigr)W,
 \qquad
 \mathfrak A_E=W^{-1}\bigl(\mathbf1\otimes\mathcal B(\mathcal H_E)\bigr)W,
 \label{eq:observablealgebras}
\end{equation}
where $\mathcal B(\mathcal H)$ denotes bounded operators. The following functionals of the material markers define the coordinates and the phase reference used for state preparation and readout. Displacements on the periodic domain are evaluated using continuous lifts of the configuration maps, fixed by continuation from the identity.

For the two-shear model, with $V_M=(2\pi\ell)^3$, define
\begin{align}
 Q[\mathbf X]&=\frac4{V_M\ell}\int_M
 \sin(a_2/\ell)\cos(a_3/\ell)(X^1(\mathbf a)-a_1)\dd^3a,
 \label{eq:markerQ}\\
 R[\mathbf X]&=\frac4{V_M\ell}\int_M
 \sin(X^1(\mathbf a)/\ell)\cos(a_2/\ell)(X^3(\mathbf a)-a_3)\dd^3a.
 \label{eq:markerR}
\end{align}
For the configurations in Eq.~\eqref{eq:two_map}, the functionals satisfy $Q[\mathbf X_{q,r}]=q$ and $R[\mathbf X_{q,r}]=r$. The material markers fix the shear origin relative to the relabeling $\mathbf X_{q+c,r}=\mathbf X_{q,r}\circ\Phi_1^c$ and define the phase reference for comparing momentum sectors.

The four-coordinate model admits an analogous local coordinate definition. Using the tangent fields at the identity in Eq.~\eqref{eq:sharedflowderivatives}, set
\begin{equation}
 M_A[\mathbf X]=\rho_0\int_M\mathbf e_A(\mathbf a)\cdot
                 [\mathbf X(\mathbf a)-\mathbf a]\dd^3a.
 \label{eq:fourmarkers}
\end{equation}
Their Jacobian at the identity is the nonsingular matrix $G_{AB}(0)$, so these functionals form local coordinates. For arbitrary $q$, choose $q_*=Q[\mathbf X]$. Boundedness of the first component of $\mathbf w_4$ implies $|q_*-q|=O(|s_4|)$ uniformly in $q$. Composing the map on the right with $\Phi_1^{-q_*}$ brings it close to the identity, so the same local coordinate construction applies in a fixed small neighborhood of $(r,s_3,s_4)=0$ for every $q\in\mathbb R$.

The wavepacket in $k=|p|$ is constructed from the normalized real even profile
\begin{equation}
 w_0(u)=\begin{cases}
 N_0^{-1/2}\exp[-1/(2(1-u^2))],&|u|<1,\\
 0,&|u|\ge1,
 \end{cases}
 \qquad N_0=\int_{-1}^1e^{-1/(1-u^2)}\dd u.
 \label{eq:bump}
\end{equation}
For $k_0>\Delta k>0$, the normalized wavepacket is
\begin{equation}
 g(k)=(\Delta k)^{-1/2}w_0((k-k_0)/\Delta k).
 \label{eq:wavepackets}
\end{equation}
Its support is $[k_0-\Delta k,k_0+\Delta k]\subset(0,\infty)$. For fixed widths, the product wavefunction in Eq.~\eqref{eq:linkedpacket} and the adapted amplitude in Eq.~\eqref{eq:adaptedstate} are smooth and compactly supported. The corresponding states have finite kinetic energy. We evaluate $C_t$ from Eq.~\eqref{eq:generalcurvatureloss} using the full kinetic metric.

\section{Numerical methods}
\label{app:numerics}
The Gaussian state in Figure~\ref{fig:kernel} has the flat-measure representation
\begin{equation}
 \psi_{\mathrm G}(r)=\mathcal N_{\mathrm G}\,a(r)^{1/4}
 \exp\!\left[-\frac{(r-r_0)^2}{4\sigma_{\mathrm G}^2}\right]\chi_L(r),
 \qquad r_0=1.3,\quad \sigma_{\mathrm G}=0.6.
 \label{eq:gaussianpreparation}
\end{equation}
Here $\mathcal N_{\mathrm G}$ normalizes the state, $a(r)^{1/4}$ accounts for the metric measure, and $\chi_L$ is a smooth cutoff compatible with homogeneous Dirichlet conditions at $r=\pm L$, with $L=12$. The conditional Hamiltonians are discretized by centered finite differences.

For equal populations of the two momentum signs, the mean energy is
\begin{equation}
 \overline E=\frac12\int_\Omega\left[
 \hbar^2G^{ab}\bigl[(\partial_a-l_a)\psi_E\bigr]^*
                         (\partial_b-l_b)\psi_E
 +\langle k^2\rangle_gG^{qq}|\psi_E|^2\right]\dd^2y,
 \label{eq:initialenergy}
\end{equation}
where $G_\vartheta$ is the induced metric, $\mu=\sqrt{\det G_\vartheta}$, and $l_a$ is defined in Eq.~\eqref{eq:generalB}. The widths are chosen to match the mean energies in Table~\ref{tab:shapedesign} before projection, with momentum parameters specified in Section~\ref{sec:quantumnumeric}. The moments of $B_\vartheta$ are evaluated directly before projection.

The covariant metric is reconstructed from the generating flows by tensor-product Chebyshev interpolation and pulled back to the three-coordinate submanifold before inversion and evaluation of half-density terms. For $d=2$, the product sine basis satisfying Dirichlet boundary conditions is
\begin{equation}
 s_j(x)=b^{-1/2}\sin\frac{j\pi(x+b)}{2b},\qquad
 \varphi_{\mathbf j}(y)=\prod_{a=1}^d s_{j_a}(y^a),\qquad
 V_N=\operatorname{span}\{\varphi_{\mathbf j}:1\le j_a\le N\}.
 \label{eq:sinebasis}
\end{equation}
For each $k$, define
\begin{equation}
 \eta_{\pm,k}(u)=\bigl(\pm ik u/\hbar,
          (\partial_1-l_1)u,\ldots,(\partial_d-l_d)u\bigr).
 \label{eq:etadef}
\end{equation}
The Hamiltonian matrix elements are evaluated by tensor-product Gauss--Legendre quadrature of the form
\begin{equation}
 \mathfrak h_{\pm,k}[u,v]=\frac{\hbar^2}{2}\int_\Omega
       \eta_{\pm,k}(u)_i^*G^{ij}\eta_{\pm,k}(v)_j\dd^dy,
 \qquad u,v\in V_N.
 \label{eq:galerkinform}
\end{equation}
Time evolution is obtained by diagonalizing the Hamiltonian matrices. The conditional states $\psi_\pm(k,t)$ evolve under $H(\pm k)$ from the same normalized projected state. We compute $f_g(t)$ from their conditional overlaps. Small losses are evaluated using
\begin{equation}
 1-|f_g|^2=\int |g(k)|^2
       \|\psi_+(k,t)-f_g(t)\psi_-(k,t)\|_2^2\dd k,
 \label{eq:stableloss}
\end{equation}
which reduces subtractive cancellation.

We use $64^2$ and $72^2$ basis functions for $b=0.08$ and $0.09$, respectively, maintaining the maximum wavenumber. The coherence-loss curves are stable under basis and quadrature refinement for the same initial states. The delay in the first crossing for the larger domain persists under refinement, and the second moments continue to cross before the coherence-loss curves. The Galerkin value of $C_\tau$ for the adapted state is about $5\%$ above its direct evaluation before basis projection.

The Euler equation is solved independently using a dealiased Fourier spectral method. The reduction in velocity error and its short-time scaling persist under spatial and temporal refinement. Numerical results for the two-shear coherence kernel are stable under grid refinement. The coefficients in Table~\ref{tab:shapedesign} remain unchanged at the quoted precision under refinement of the metric interpolation and quadrature. The interpolated metric is positive definite throughout the computational domain. Numerical code and data are supplied as supplementary material.
\FloatBarrier
\section{Convergence of the reduced quantum dynamics}
\label{app:channelproof}
We establish convergence of $C_t$ and the logical dephasing channel under basis refinement and metric approximation. For the three-coordinate model, consider $\Omega=(-b,b)^2$, a smooth metric uniformly positive on a neighborhood of $\overline\Omega$, and normalized $g\in C_c^\infty(\mathbb R_+)$. The continuum quadratic forms corresponding to Eq.~\eqref{eq:galerkinform} have common domain $H_0^1(\Omega)$ in the half-density representation.

Let $P_N$ project onto the sine space in Eq.~\eqref{eq:sinebasis}, set $\psi_{E,N}=P_N\psi_E/\|P_N\psi_E\|_2$, and write $B_N=P_NB|_{V_N}$. For normalized $\psi_E\in H_0^1(\Omega)$, the sine expansion converges in $H^1$. Smooth coefficients make $B:H_0^1(\Omega)\to L^2(\Omega)$ bounded, with a constant $K_B$ such that
\begin{equation}
 \|B_N\psi_{E,N}-B\psi_E\|_2
 \le K_B\|\psi_{E,N}-\psi_E\|_{H^1}
       +\|(P_N-1)B\psi_E\|_2\longrightarrow0.
 \label{eq:Bestimate}
\end{equation}
The first and second moments of $B$ therefore converge. Continuity of the energy form in $H^1$, uniformly over the compact support of $g$, implies convergence of the mean energy and $C_t$. The expansion in time requires the moment condition in Appendix~\ref{app:geometry}.

For finite-time evolution, take any normalized $\psi_E\in L^2(\Omega)$. Let $H_{\pm,N}(k)$ be the Hamiltonian associated with the restriction of the continuum form to $V_N$ in Eq.~\eqref{eq:galerkinform}, and define
\begin{equation}
 f_N(t)=\int |g(k)|^2
 \left\langle e^{-itH_{-,N}(k)/\hbar}\psi_{E,N},
               e^{-itH_{+,N}(k)/\hbar}\psi_{E,N}\right\rangle\dd k.
 \label{eq:finiteNoverlap}
\end{equation}
The exact overlap $f_g$ is given by Eq.~\eqref{eq:generalcurvatureloss}. At fixed $k$, the shifted forms are continuous and coercive on $H_0^1(\Omega)$, and the nested sine spaces are dense there. The Galerkin resolvents converge strongly, and the associated unitary groups converge strongly and uniformly on compact time intervals~\cite{kato}. The overlap difference is bounded by two, so dominated convergence under the $|g|^2$ integral gives uniform convergence of $f_N$.

For the logical dephasing channel in Eq.~\eqref{eq:logicalchannelmatrix},
\begin{equation}
 \|\mathcal D_{f_1}-\mathcal D_{f_2}\|_\diamond=|f_1-f_2|.
 \label{eq:diamondidentity}
\end{equation}
An output phase rotation reduces the difference to $|f_1-f_2|(\id-Z_{\mathrm L}\,\cdot\,Z_{\mathrm L})/2$. Its diamond norm is at most $|f_1-f_2|$, attained on $(\ket0+\ket1)/\sqrt2$. Therefore, for every $T<\infty$,
\begin{equation}
 \sup_{|t|\le T}|f_N(t)-f_g(t)|\longrightarrow0,
 \qquad
 \sup_{|t|\le T}\|\mathcal D_{f_N(t)}-\mathcal D_{f_g(t)}\|_\diamond
       \longrightarrow0.
 \label{eq:finiteoverlapconvergence}
\end{equation}
Metric reconstruction is controlled separately. If $G_j\to G$ in $W^{1,\infty}(\Omega)$ with a common positive lower bound, the inverse metrics and half-density coefficients converge uniformly. For $k$ in the support of $g$,
\begin{equation}
 |\mathfrak h^{(j)}_{\pm,k}[u,v]-\mathfrak h_{\pm,k}[u,v]|
 \le\epsilon_j\|u\|_{H^1}\|v\|_{H^1},\qquad
 \|B_j-B\|_{H^1\to L^2}\longrightarrow0,
 \quad\epsilon_j\longrightarrow0.
 \label{eq:metricformstability}
\end{equation}
The same resolvent argument establishes convergence of the conditional evolutions and logical dephasing channel. For fixed $\psi_E\in H_0^1(\Omega)$, the mean energy and $C_t$ also converge under metric approximation followed by basis refinement.

\end{document}